\documentstyle[11pt]{article}
\normalbaselines
\newcounter{eqnletter}[equation]
\catcode`\@=11
\@addtoreset{equation}{section}   
\catcode`\@=12

\begin{document}

\begin{center}

{\LARGE\bf Non-Linear Quantization of \\[5mm]
Integrable Classical Systems}

\vskip 1cm

{\large {\bf Antonio Scotti} }\footnote{This work was
partially supported by Istituto Nazionale di Fisica
Nucleare I.N.F.N. and Istituto Nazionale di Fisica
della Materia I.N.F.M.}

\vskip 0.1 cm

Dipartimento di Fisica, Universita di Parma,\\
Viale delle Scienze, I - 43100 Parma, Italy \footnote{E-mail
address: scotti@parma.infn.it} \\[0.5cm]

and\\[0.5cm]

{\large {\bf Alexander Ushveridze} }\footnote{This work was
partially supported by KBN grant no. 2P30221706p01}

\vskip 0.1 cm

Department of Theoretical Physics, University of Lodz,\\
Pomorska 149/153, 90-236 Lodz, Poland\footnote{E-mail
address: alexush@parma.infn.it and alexush@krysia.uni.lodz.pl} \\

\end{center}
\vspace{1 cm}
\begin{abstract}

It is demonstrated that the so-called ``unavoidable quantum
anomalies'' can be avoided in the framework of a special non-linear
quantization scheme. In this scheme, the quantized hamiltonians
are represented by non-linear but homogeneous operators in 
Hilbert space. The  nonlinear terms are of the same order as 
quantum anomalies, and their role is to cancel anomalies. 
The quantization method proposed is applicable to integrable
classical dynamical systems and the result of 
quantization is again an integrable (but, generally, non-linear) 
"quantum" system. A simple example is discussed in detail. 
Irrespective of the existence of possible physical applications,
the method provides a constructive way for 
extending the notion of quantum integrability to non-linear 
spectral problems and gives a practical tool for buiding completely 
integrable non-linear spectral equations in Hilbert space.

\end{abstract}

\newpage 

\section{Introduction}
\label{in}

It is known that quantization does not generally preserve 
zero commutation relations between observables. In other words,
classically integrable systems may cease to be integrable
after their quantization. Of course, the quantization
procedure is not unique. There are infinitely
many ways to  quantize a system and, 
for this reason, the integrability can sometimes 
be recovered after an appropriate choice of a
quantization scheme. However, the typical situation
(especially for multi-dimensional systems) is that 
there are no quantization schemes at all in which such a
recovering would be possible. In such a case we speak of the ``unavoidable
quantum anomalies''. 

In order to understand the reason for the existence of
unavoidable quantum anomalies, consider a certain
completely integrable $N$-dimensional classical system with
$N$ mutually commuting integrals of motion $H_n, \
n=1,\ldots, N$. Denote by $\hat H_n^{(0)}, \
n=1,\ldots, N$ the quantum versions of these classical integrals
of motion obtained within some fixed quantization scheme.
Assume that operators $\hat H_n^{(0)}$ do not commute with each other. For a
randomly chosen quantization scheme this is a typical situation.
Computing the commutator of these operators, we obtain
\begin{eqnarray}
[\hat H_n^{(0)},\hat H_m^{(0)}] = \hbar^2 \hat F_{nm}^{(1)}+\hbar^3
\hat F_{nm}^{(2)}+\ldots.
\label{in.1}
\end{eqnarray}
Obviously, this expansion should start with
$\hbar^2$. Otherwise we would obtain non-zero
Poisson brackets in the classical limit, which is impossible
because $H_n, \ n=1,\ldots, N$ are, by assumption, integrals of motion.
Now remember that the chosen quantization scheme is not unique.
This is a trivial consequence of the fact that, instead
of operators $\hat H_n^{(0)}, \ n=1,\ldots, N$, we could
consider arbitrary operators of the form
\begin{eqnarray}
\hat H_n=\hat H_n^{(0)}+\hbar \hat H_n^{(1)}+\hbar^2 \hat H_n^{(2)}+\ldots,
\qquad n=1,\ldots, N.
\label{in.2}
\end{eqnarray}
which, evidently, would have the same classical limit. This
fact suggests to improve the commutation relations (\ref{in.1})
by replacing the operators $\hat H_n^{(0)}$ by more general operators
(\ref{in.2}) with appropriately chosen corrections 
$\hat H_n^{(1)}, \hat H_n^{(2)},\ldots$. Requiring that $[\hat H_n,
\hat H_m]=0$,  using (\ref{in.1}) and (\ref{in.2}) and
collecting the terms of the same order in $\hbar$, we
obtain the system of recurrence relations for corrections
of our interest. For example, the equation determining the 
first correction reads
\begin{eqnarray}
[\hat H_n^{(0)},\hat H_m^{(1)}]+[\hat H_n^{(1)},\hat
H_m^{(0)}]+\hbar \hat F_{nm}^{(1)}=0, \qquad n>m,
\qquad n,m=1,\ldots, N.
\label{in.3}
\end{eqnarray}
However, even this first equation clearly demonstrates that
the solution of the problem does not generally exist. 
Indeed, the number of unknowns in equation (\ref{in.3}) is $N$.
At the same time, the number of equations is $N(N-1)/2$. If
$N>3$, i.e., for more than three dimensional models, the
system becomes overdetermined and generally has no
solutions. This is what we meant saying that for multi-dimensional
models the situation with quantum anomalies is 
typical and cannot be avoided in the framework of standard
formalizm of quantum 
mechanics\footnote{Of course, in the presence of some 
symmetry in a model, the equations for corrections may
cease to be independent and their effective number may decrease. 
In this case the quantization
without anomalies may become possible. Such situation is
usually realized in models obtainable in the framework of
the r-matrix method if the underlying symmetry is $su(2)$
or $su(3)$. As much as we know, for higher symmetries the problem with
quantum anomalies in r-matrix approach remains still unsolved.}.

The aim of the present paper is to demonstrate that
construction of quantization schemes free of any quantum
anomalies (at least on the states) 
becomes possible if one drops out the condition
of linearity of quantum mechanics. The resulting non-linear
Schr\"odinger equations are very simalar to Doebner --
Goldin equations \cite{dob1} recently derived in the
framework of a rather general quantization scheme based on the
use of infinite-dimensional algebras of vector fields and
group of diffeomorphisms \cite{gol,ang}.
Speaking of non-linear Schr\"odinger equations and
non-linear quantum mechanics, we mean a
hypothetic quantum theory in which the observables are
represented by non-linear operators in Hilbert space. 
At the present time there are many examples of 
such theories (see e.g. refs. \cite{bial,kib,schu,wein,dob1} 
and references therein). 
Because up to now there are no experemental
indications that the quantum world is non-linear, 
the measure of non-linear effects in all reasonable theories
should be very small. In our scheme this condition is automatically
satisfied because in this scheme the role of
non-linear terms is to cancel quantum anomalies, and
therefore, the order of these terms is the same as the
order of anomalies, i.e., at most $\hbar^2$.

In this paper we restrict ourselves to discussing a 
particular (but very important) case of quantization of
completely integrable classical systems. We leave the consideration
of the general case to later publications.
In the case of completely integrable systems, the procedure
of quantization  consists of three steps:
1) separation of variables in an integrable classical 
system\footnote{The separation of variables is understood
here in the generalized (Sklyanin) sense. Following
Sklyanin \cite{skl}, 
we call a $N$-dimensional classical system separable if there
exist such a {\it canonical transformation} to canonically conjugated
variables $p_i,q_i,\ i=1,\ldots, N$ in which the Hamilton --
Jacobi equation for the system becomes equivalent to a
system of one-dimensional multi-parameter spectral equations
$F_i(q_i,p_i,h_1,\ldots,h_N),\ i=1,\ldots,N$ with some polynomial
functions $F_i$. For more details see ref. \cite{skl}.},
2) standard quantization of one-dimensional equations so obtained, and 
3) reconstruction of an integrable quantum system from the quantized
one-dimensional equations. 

The first step is most non-trivial from the practical point of view,
because, in practice, it is often very difficult to check whether a given
classical system is integrable or not, and even if it is, in
which canonical coordinates it is separable. However,
this difficulty can be avoided by using the ``classical inverse
method of separation of variables'' (classical IMSV), 
when instead of checking the integrability or
separability of a given system, we are
simply building systems which are separable and integrable
by construction \cite{skl,dobush}. 

The last step is nothing else than the so-called ``quantum
inverse method of separation of variables'' (quantum IMSV),
which, up to now, has been applied to systems of separated 
equations with linear
dependence on separation constants (see e.g. refs. 
\cite{dobush,skl2,ushleb,ushbook}). 
For such equations the resulting
quantum systems are always linear. The non-linear quantum operators
appear only if the separated equations depend on separation
constants non-linearly. 

These two classical and quantum versions of IMSV will 
be discussed in detail in the two subsequent sections \ref{cl}
and \ref{qu}. In section \ref{no} we explain what we mean
of non-linear quantization. The last section \ref{ex} is devoted
to  discussion of specific two-dimensional models for which
all the aspects of the proposed quantization scheme become
transparent and clear. Unfortunately, these models, because
of their two-dimensionality, are not the examples of models with true 
quantum anomalies. The discussion of the latter would force
us to work in at least four-dimensional space, which, of course
would lead to very cumbersome expressions, non-desirable in
the framework of this paper.

\section{Classical version of IMSV}
\label{cl}

Let $F_i=F_i(x,y,z_1,\ldots, z_N),\ i=1,\ldots, N$ be $N$
(may be partially or completely coinciding) polynomial
functions of $N+2$ complex variables $x,y,z_1,\ldots,z_N$.
Consider $N$ ordinary first-order differential equations
\begin{eqnarray}
F_i\left(q_i,\frac{\partial S_i(q_i)}{\partial q_i},h_1,\ldots,h_N\right)=0,
\quad i=1,\ldots, N
\label{a.1}
\end{eqnarray} 
for $N$ functions $S_i(q_i),\ i=1,\ldots, N$. Obviously, all these equations
can be explicitly integrated after solving the algebraic equations
$F_i(x,y,z_1,\ldots,z_N)=0$ with respect to the second
variable, $y=Y_i(x,z_1,\ldots,z_N)$. 
We shall write the result in the following form
\begin{eqnarray}
S_i(q_i)=\int Y_i(q_i,h_1,\ldots,h_N)\mbox{d}q_i, \quad i=1,\ldots, N
\label{a.2}
\end{eqnarray}
stressing the fact that all solutions depend on $N$
arbitrary numbers $h_1,\ldots,h_N$ parametrizing equations (\ref{a.1}).

Let us now note that the equalities (\ref{a.1}) are not
violated if we replace the functions $S_i(q_i)$ in
them by their sum
\begin{eqnarray}
S({\bf q})=\sum_{k=1}^N S_k(q_k).
\label{a.3}
\end{eqnarray} 
This gives us the following new system of equations
\begin{eqnarray}
F_i\left(q_i,\frac{\partial S({\bf q})}
{\partial q_i}, h_1,\ldots,h_N\right)=0,
\quad i=1,\ldots, N,
\label{a.4}
\end{eqnarray}
for a single function $S({\bf q})$ of $N$ variables $q_1,\ldots,q_N$.
This system admits (by construction) a total separation of
variables and its solution is given by formulas (\ref{a.3})
and (\ref{a.2}). 

At the same time, the system (\ref{a.4}) 
can be interpreted as a system of $N$ algebraic equations for $N$
unknown variables $h_1,\ldots, h_N$.
Solving it with respect to $h_1,\ldots, h_N$ we obtain
\begin{eqnarray}
h_\alpha=H_\alpha\left({\bf q},\frac{\partial S({\bf
q})}{\partial {\bf q}}\right),
\quad \alpha=1,\ldots, N,
\label{a.5}
\end{eqnarray}
where $H_\alpha({\bf q},{\bf p})$ are some functions of two
$N$-component vector variables ${\bf q}=\{q_1,\ldots,q_N\}$ and
${\bf p}=\{p_1,\ldots,p_N\}$. Interpreting these variables as
canonically conjugated coordinates and momenta, $\{q_\alpha,
p_\beta\}=\delta_{\alpha\beta}$, let us consider a
$N$-dimensional classical dynamical system with the hamiltonian
\begin{eqnarray}
H({\bf q},{\bf p})=E(H_1({\bf q},{\bf p}),\ldots, H_r({\bf
q},{\bf p}))
\label{a.6}
\end{eqnarray}
in which $E(h_1,\ldots,h_N)$ is an arbitrarily fixed 
function of $h_1,\ldots, h_N$.

It is not difficult to see that the stationary Hamilton--Jacobi
equation for the hamiltonian $H({\bf q},{\bf p})$,
\begin{eqnarray}
H\left({\bf q},\frac{\partial S({\bf
q})}{\partial {\bf q}}\right)=E,
\label{a.7}
\end{eqnarray}
is separable by construction and its complete solution
(i.e. a solution parametrized by $N$
arbitrary parameters) has the form
\begin{eqnarray}
E=E(h_1,\ldots,h_N), \quad S({\bf q})=\sum_{i=1}^N
\int Y_i(q_i,h_1,\ldots,h_N)\mbox{d}q_i. 
\label{a.8}
\end{eqnarray} 
This immediately follows from the fact that, after fixing the form of
the function $E(h_1,\ldots,h_N)$, the equation (\ref{a.7}) becomes equivalent
to the separable system (\ref{a.6}).

So we have demonstrated that any set of $N$ functions 
$F_i(x,y,z_1,\ldots,z_N)$, $i=1,\ldots N$ generates a certain separable
classical dynamical system by means of the procedure which we shall 
refer below to as the {\it classical 
inverse procedure of separation of variables}.

It is not difficult to show that the classical system obtained in 
such a way is not only separable but also completely integrable in the
sense that its hamiltonian admits enough mutually commuting 
integrals of motion. In order to demonstrate this fact, it
is sufficient to show that functions 
$H_\alpha({\bf q},{\bf p}),\ \alpha=1,\ldots, N$
form a commutative family.

Indeed, if $H_\alpha=H_\alpha({\bf q},{\bf p}),
\alpha=1,\ldots N$, then we have by definition
\begin{eqnarray}
F_i=F_i(q_i,p_i,H_1,\ldots,H_N)=0,
\quad i=1,\ldots, N.
\label{b.1}
\end{eqnarray} 
Therefore
\begin{eqnarray}
\frac{d F_i}{d q_k}=\frac{\partial F_i}{\partial q_k}+
\sum_{l=1}^N\frac{\partial F_\alpha}{\partial H_l}\frac{\partial
H_l}{\partial q_k}=0,
\label{b.2a}
\end{eqnarray}  
\begin{eqnarray}
\frac{d F_i}{d p_k}=\frac{\partial F_i}{\partial p_k}+
\sum_{l=1}^N\frac{\partial F_i}{\partial H_l}\frac{\partial
H_l}{\partial p_k}=0,
\label{b.2b}
\end{eqnarray}
and, consequently,
\begin{eqnarray}
\sum_{l=1}^N\frac{\partial F_i}{\partial H_l}\frac{\partial
H_l}{\partial q_k}=-\frac{\partial F_i}{\partial q_k},
\label{b.3a}
\end{eqnarray}  
\begin{eqnarray}
\sum_{l=1}^N\frac{\partial F_i}{\partial H_l}\frac{\partial
H_l}{\partial p_k}=-\frac{\partial F_i}{\partial p_k}.
\label{b.3b}
\end{eqnarray}
On the other hand, we have
\begin{eqnarray}
0=\{F_{i_1},F_{i_2}\}=\sum_{k=1}^N\left[
\frac{d F_{i_1}}{d q_k}\frac{d F_{i_2}}{d p_k}-
\frac{d F_{i_1}}{d p_k}\frac{d F_{i_2}}{d q_k}\right]
=\nonumber\\
=\sum_{k=1}^N\left[
\frac{\partial F_{i_1}}
{\partial q_k}
\frac{\partial F_{i_2}}
{\partial p_k}-
\frac{\partial F_{i_1}}
{\partial p_k}
\frac{\partial F_{i_2}}
{\partial q_k}\right]+\nonumber\\
+
\sum_{k,l=1}^N\left[
\frac{\partial F_{i_1}}
{\partial q_k}
\frac{\partial F_{i_2}}
{\partial H_l}
\frac{\partial H_l}
{\partial p_k}-
\frac{\partial F_{i_1}}
{\partial p_k}
\frac{\partial F_{i_2}}
{\partial H_l}
\frac{\partial H_l}
{\partial q_k}\right]+\nonumber\\
+
\sum_{k,l=1}^N\left[
\frac{\partial F_{i_1}}
{\partial H_l}
\frac{\partial H_l}
{\partial q_k}
\frac{\partial F_{i_2}}
{\partial p_k}-
\frac{\partial F_{i_1}}
{\partial H_l}
\frac{\partial H_l}
{\partial p_k}
\frac{\partial F_{i_2}}
{\partial q_k}\right]+\nonumber\\
+
\sum_{k,l=1}^N
\frac{\partial F_{i_1}}
{\partial H_k}
\frac{\partial F_{i_2}}
{\partial H_l}
\{H_k,H_l\}.
\label{b.4}
\end{eqnarray} 
Applying (\ref{b.3a}) and (\ref{b.3b}) to (\ref{b.4}) we obtain
\begin{eqnarray}
\sum_{k,l=1}^N
\frac{\partial F_{i_1}}
{\partial H_k}
\frac{\partial F_{i_2}}
{\partial H_l}
\{H_k,H_l\}=
\sum_{k=1}^N\left[
\frac{\partial F_{i_1}}
{\partial q_k}
\frac{\partial F_{i_2}}
{\partial p_k}-
\frac{\partial F_{i_1}}
{\partial p_k}
\frac{\partial F_{i_2}}
{\partial q_k}\right].
\label{b.5}
\end{eqnarray}
Since the functions $F_{i_1}$ and $F_{i_2}$ depend
on different variables if $i_1 \neq i_2$, we have
\begin{eqnarray}
\sum_{k,l=1}^N
\frac{\partial F_{i_1}}
{\partial H_k}
\frac{\partial F_{i_2}}
{\partial H_l}
\{H_k,H_l\}=0,
\label{b.6}
\end{eqnarray}
for all $i_1$ and $i_2$, and, because of the
invertibility of the matrix $||\partial F_i/
\partial H_k ||$, we have
\begin{eqnarray}
\{H_k,H_l\}=0.
\label{b.7}
\end{eqnarray}
It is not difficult to see that the functions $H_1,\ldots,H_N$
constructed in such a way are functinally independent and thus
the classical system with hamiltonian $H=E(H_1,\ldots,H_N)$ is
completely integrable. 

\section{Quantum version of IMSV}
\label{qu}

Let $F_i=F_i(x,y,z_1,\ldots, z_N),\ i=1,\ldots, N$ be now $N$
(may be partially or completely coinciding) polynomial 
functions of $N+2$ generally non-commuting 
variables $x,y,z_1,\ldots,z_N$. We assume that these variables 
are distributed in the expression in that order in which
they are written. Consider $N$ linear differential equations
\begin{eqnarray}
F_i\left(q_i,\mbox{i}
\hbar\frac{\partial}{\partial q_i},h_1,\ldots,h_N\right)\Psi_i(q_i)=0,
\quad i=1,\ldots, N
\label{c.1}
\end{eqnarray} 
for $N$ functions $\Psi_i(q_i), i=1,\ldots, N$. We shall
call equations of the type (\ref{c.1}) {\it multi-parameter
spectral equations}. The role of the spectral parameters in
them is played by the numbers $h_1,\ldots,h_N$. The problem is to
find all admissible values of these parameters for which
the system (\ref{c.1}) has solutions belonging to a certain {\it
a priori} given classes $W_i$ of functions $\Psi_i(q_i)\in W_i,\ 
i=1,\ldots, N$.
The set of all admissible ``$N$-plets'' $\{h_1,\ldots, h_N\}$ will
be called the {\it spectrum} of the system (\ref{c.1}).
Below we shall assume that the classes $W_i,\ i=1,\ldots,N$ 
are chosen in such a way that the system (\ref{c.1}) has a discrete spectrum.
The corresponding discrete set of solutions we represent as
\begin{eqnarray}
\Psi_i(q_i)=\Xi_i(q_i,h_1,\ldots,h_N), \quad i=1,\ldots, N,
\label{c.2}
\end{eqnarray}
stressing their correspondence to the values of spectral
parameters $h_1,\ldots, h_N$.

Let us now note that the equalities (\ref{c.2}) are not
violated if we replace the functions $\Psi_i(q_i)$ in
them by their product
\begin{eqnarray}
\Psi({\bf q})=\prod_{i=1}^N \Psi_\alpha(q_i)
\label{c.3}
\end{eqnarray} 
belonging to the class $W=\otimes_{i=1}^N W_i$.
This gives us the following new system of equations
\begin{eqnarray}
F_i\left(q_i,\mbox{i}\hbar\frac{\partial}
{\partial q_i}, h_1,\ldots,h_N\right)\Psi({\bf q})=0,
\quad i=1,\ldots, N,
\label{c.4}
\end{eqnarray}
for a single function $\Psi({\bf q})$ of the $N$-component
vector variable ${\bf q}=\{q_1,\ldots,q_N\}$.
This system admits (by construction) a total separation of
variables and its solution is given by formulas (\ref{c.3})
and (\ref{c.2}). 

At the same time, the system (\ref{c.4}) 
can be interpreted as a system of $N$ algebraic equations for $N$
unknown spectral parameters $h_1,\ldots, h_N$.
Solving it with respect to $h_1,\ldots, h_N$ and
multiplying the result by $\Psi({\bf q})$ we obtain
\begin{eqnarray}
h_\alpha\Psi({\bf q})=\hat H_\alpha\left({\bf q},\mbox{i}\hbar\frac{\partial}
{\partial {\bf q}}\right)\Psi({\bf q}),
\quad \alpha=1,\ldots, N,
\label{c.5}
\end{eqnarray}
where $\hat H_\alpha({\bf q},{\bf p})$ 
is some formal writing for, generally, non-linear differential operators
built from two $N$-component non-commuting operators  
${\bf q}=\{q_1,\ldots,q_N\}$ and 
${\bf p}=\{\mbox{i}\hbar\partial/\partial q_1,\ldots,
\mbox{i}\hbar\partial/\partial q_N\}$. We interpret these operators as
operators of coordinates and momenta satisfying the Heisenberg
commutation relations $[{\bf q}\otimes{\bf p}]=\mbox{i}\hbar{\bf I}$. 

Consider the $N$-dimensional non-linear operator
\begin{eqnarray}
\hat H({\bf q},{\bf p})=
E(\hat H_1({\bf q},{\bf p}),\ldots, \hat H_r({\bf q},{\bf p}))
\label{c.6}
\end{eqnarray}
where $E(h_1,\ldots,h_N)$ is the same
polynomial function of $N$ variables $h_1,\ldots, h_N$ as in
the previous section. At this point we only note that if the
arguments of this function are non-commuting variables,
they should be distributed in the expression in that order
in which they are written. The product of non-linear
operators will  be understood hereafter as their composition.

It is easy to see that the non-linear spectral equation 
\begin{eqnarray}
\hat H\left({\bf q},\mbox{i}\hbar\frac{\partial}{\partial {\bf
q}}\right)\Psi({\bf q})=E\Psi({\bf q}),
\label{c.7}
\end{eqnarray}
is separable by construction and its complete solution
(i.e. a solution parametrized by $N$ spectral parameters) has the form
\begin{eqnarray}
E=E(h_1,\ldots,h_N), \quad \Psi({\bf q})=\prod_{\alpha=1}^N
\Xi_{i[\alpha]}(q_\alpha,h_1,\ldots,h_N). 
\label{c.8}
\end{eqnarray} 
This immediately follows from the fact that, after fixing the form of
the function $E(h_1,\ldots,h_N)$, the equation (\ref{c.7}) becomes equivalent
to the separable system (\ref{c.4}).

So, we have demonstrated that any $N$ arbitrarily chosen functions 
$F_i(x,y,z_1,\ldots,z_N)$, $r=1,\ldots N$ generate a
certain, in general, non-linear separable
spectral equation by means of the 
procedure which we shall refer to as the {\it quantum 
inverse procedure of separation of variables}. 
The equation (\ref{c.7}) will be called the (non-linear) Schr\"odinger
equation. 

Now we show that the operators $\hat H_{\alpha}=\hat H_{\alpha}
({\bf q},{\bf p})$ can (in some sense) be considered as
integrals of motion of a certain completely integrable non-linear
"quantum" system. Indeed, from the definition of these operators  
it immediately follows that they are homogeneous operators of order one, i.e.,
for any constant $c$
\begin{eqnarray}
\hat H_{\alpha}(c\Psi)=c\hat H_{\alpha}\Psi.
\label{c.9}
\end{eqnarray} 
Moreover, as follows from (\ref{c.5}), these operators have
a common set of eigenfunctions:
\begin{eqnarray}
\hat H_{\alpha}\Psi=h_{\alpha}\Psi, \quad \alpha=1,\ldots, N.
\label{c.10}
\end{eqnarray} 
By using these two properties, we can consider two chains of equalities:
\begin{eqnarray}
\hat H_{\alpha}\hat H_{\beta}\Psi=\hat H_{\alpha} h_{\beta}\Psi =
h_{\beta}\hat H_{\alpha}\Psi=h_{\beta}h_{\alpha}\Psi
\label{c.11a}
\end{eqnarray} 
and
\begin{eqnarray}
\hat H_{\beta}\hat H_{\alpha}\Psi=\hat H_{\beta} h_{\alpha}\Psi =
h_{\alpha}\hat H_{\beta}\Psi=h_{\alpha}h_{\beta}\Psi.
\label{c.11b}
\end{eqnarray} 
Subtracting (\ref{c.11b}) from (\ref{c.11a}), we find that
\begin{eqnarray}
[\hat H_{\alpha},\hat H_{\beta}]\Psi=0.
\label{c.12}
\end{eqnarray} 
Note that the operators $\hat H_\alpha$ are  
non-linear only if the multi-parameter spectral equations
(\ref{c.2}) depend on their spectral parameters non-linearly. In the
case of linear dependence on spectral parameters the
operators $\hat H_\alpha$ will obviously be linear. 
Assume that the set of solutions of equation (\ref{c.7}) is
complete in the sense that $W$ (see above) is a Hilbert space and 
any function from $W$ can be expanded in solutions of equation (\ref{c.7}). If
the operators $\hat H_\alpha$ are linear, then we can claim
that equality (\ref{c.12}) holds for all elements of $W$
and thus the operators $\hat H_\alpha$ commute in the strong
operator sense. In the case of non-linear operators
$\hat H_\alpha$ this reasoning does not work, 
and the commutativity should be understood in the in the weak sense, i.e.
on the solutions of the spectral problem (\ref{c.7}).

\section{The non-linear quantization method}
\label{no}

In the previous section we used the term ``Schr\"odinger equation'' and the
adjective ``quantum'' in order to stress the
fact that the equations (\ref{a.7}) and (\ref{c.7}) are related
to each other by some ``quantization procedure''. This
means that taking in the ``quantum Schr\"odinger equation''
(\ref{c.7}) the classical limit $\hbar\rightarrow 0$, we obtain the classical 
Hamilton--Jacobi equation (\ref{a.7}). In order to see
this, it is sufficient to represent the ``wavefunctions''
$\Psi({\bf q})$ in the form
\begin{eqnarray}
\Psi({\bf q})=\exp\left(\frac{\mbox{i}}{\hbar}S({\bf q}) \right),
\label{c.13}
\end{eqnarray} 
after which it becomes clear that the factorizability of the
wavefunctions implies the decomposability of their
logarithms, i.e. classical actions. This enables one to write
\begin{eqnarray}
\Psi_i(q_i)=\exp\left(\frac{\mbox{i}}{\hbar}
S_i(q_i) \right),
\quad i=1,\ldots,N
\label{c.14}
\end{eqnarray} 
which reduces the problem of verifying our assertion 
to checking that in the classical limit 
the multi-parameter spectral equations (\ref{c.1}) reduce to
the equations (\ref{a.1}). But this is obvious because the
substitution of (\ref{c.14}) into (\ref{c.1}) gives an
expression whose constant term in $\hbar$ exactly
coincides with (\ref{a.1}). All other terms of higher
orders in $\hbar$ vanish in the limit $\hbar\rightarrow 0$.

\section{An example}
\label{ex}

In this section we consider a simple example demonstrating
how does the proposed scheme work.
We start with two $F$-functions of the form
\begin{eqnarray}
F_1(x,y,h,g)=x^2+y^2-\frac{\epsilon}{2a^4} h^2-
\left(\frac{2}{a^2}x^2 +1\right)h+g
\label{ex.1}
\end{eqnarray}
and
\begin{eqnarray}
F_2(x,y,h,g)=x^2+y^2-\frac{\epsilon}{2a^4} h^2-
\left(\frac{2}{a^2}x^2+1\right)h-g.
\label{ex.2}
\end{eqnarray}
in which the variables $x$ and $y$ are associated with
classical or quantum coordinates and momenta and $h, g$ are
spectral parameters. We see that, generally, the functions
(\ref{ex.1}) and (\ref{ex.2}) depend on the spectral parameters non-linearly,
but if $\epsilon=0$ or $a=\infty$, this dependence becomes linear. For
this reason $\epsilon$ and $a$ will play the role of non-linear
deformation parameters of a system. 
Let us now construct classical and quantum models
associated with functions (\ref{ex.1}) and (\ref{ex.2}).

\subsection{Classical case}

According to the results of section \ref{cl}, functions
(\ref{ex.1}) and (\ref{ex.2}) can be used for building the
integrals of motion for a certain two-dimensional 
completely integrable classical dynamical system.
Denoting these integrals by $H=H({\bf q}, {\bf p})$ and
$G=G({\bf q}, {\bf p})$, where ${\bf q}=\{q_1,q_2\}$ and
${\bf p}=\{p_1,p_2\}$, and following the general
prescriptions of section \ref{cl}, we can write for them the 
following two elementary solvable equations
\begin{eqnarray}
p_1^2+q_1^2-\frac{\epsilon}{2a^4} H^2-
\left(\frac{2}{a^2}q_1^2 +1\right)H+G=0
\label{ex.3}
\end{eqnarray}
and
\begin{eqnarray}
p_2^2+q_2^2-\frac{\epsilon}{2a^4} H^2-
\left(\frac{2}{a^2}q_2^2+1\right)H-G=0.
\label{ex.4}
\end{eqnarray}
Let us take the first integral $H$ as the hamiltonian of a system. 
Define for this hamiltonian the {\it potential} 
$V({\bf q})=H({\bf q},0)$.
Despite the fact that the hamiltonian cannot be generally
represented as the sum of kinetic and potential energies,
this function still reflects some general spectral properties of the model. 
Indeed, as it follows from equations (\ref{ex.2}) and (\ref{ex.3}),
the hamiltonian $H=H({\bf q}, {\bf p})$ is an
increasing function of ${\bf p}^2$. Therefore, we have the inequality
$H({\bf q},{\bf p})\ge V({\bf q})$ which means that 
the volume of the phase space bounded by the level surface 
$H({\bf q},{\bf p})=h$ should be finite
if $h\le \max V({\bf q})$ and infinite if $h\ge \max V({\bf q})$.

In the particular case, when the parameter $a^4$ is very large,
the system (\ref{ex.3}) -- (\ref{ex.4}) becomes linear,
and, solving it, we obtain the model with hamiltonian
\begin{eqnarray}
H=\frac{{\bf p}^2+{\bf q}^2}{2},
\label{ex.7}
\end{eqnarray}
which is nothing else than the two-dimensional symmetric harmonic oscillator,
i.e., the sum of hamiltonians of one-dimensional harmonic oscillators.
The second integral of motion, G, reduces in this
case to the difference of hamiltonians of one-dimensional
harmonic oscillators, and thus, the commutativity of $H$
and $G$ is obvious. The potential $V({\bf q})={\bf q}^2/2$
of this model does not have an upper bound and the motion
in this system is always finite. 

If $a$ is finite but $\epsilon$ is small, then we still
have a linear system of equations for $H$ and $G$, whose
solution leads to a little bit more comlicated model with hamiltonian
\begin{eqnarray}
H=\frac{a^2}{2}\frac{{\bf p}^2+{\bf q}^2}{a^2+{\bf q}^2}.
\label{ex.8}
\end{eqnarray}
Although the commutativity of $H$ with the corresponding
$G$ is not so obvious as in the previous case, it also can be 
checked by direct computation of the Poisson bracket.
Now the potential of the model is bounded by $\max V({\bf q})=a^2/2$,
and therefore the model (\ref{ex.8}) describes finite motion if $h<a^2/2$ and
infinite motion if $h>a^2/2$.

In the most general case, when both $\epsilon$ and $a$ are
finite numbers, the equations  (\ref{ex.3}) -- (\ref{ex.4})
become quadratic and have now two different solutions for
$H$. For definiteness we choose that solution which is
continuously connected with particular solutions (\ref{ex.7}) and 
(\ref{ex.8}). It has the form
\begin{eqnarray}
H=\frac{a^2}{\epsilon^2}({\bf q}^2+a^2)\left\{-1
+\sqrt{1+\frac{\epsilon^2({\bf p}^2+{\bf q}^2)}{(a^2+{\bf q}^2)^2}}\right\}.
\label{ex.5}
\end{eqnarray}
Reconstructing the second solution $G$ of this system, and
checking its commutativity with $H$ we can again make sure
that the obtained model is completely integrable.
As before, its potential is monotonically increasing
function of ${\bf q}^2$ and tends to $\max V({\bf q})=a^2/2$
if ${\bf q}^2\rightarrow\infty$.

Despite the fact that the model (\ref{ex.5}) looks
rather complicated, its Hamilton -- Jacobi equation 
is separable by substitution $S({\bf q})=S_1(q_1)+S_2(q)$, 
and reduces to the system of two one-dimensional equations
for $S_1(q_1)$ and $S_2(q_2)$. This system is nothing else than
the initial system (\ref{ex.3}) -- (\ref{ex.4}) with
$p_1=\partial S_1(q_1)/\partial q_1$, $p_2=\partial S_2(q_2)/\partial q_2$,
$H=h$ and $G=g$. Solving it, we obtain the complete integral
of the model
\begin{eqnarray}
S({\bf q})= \int dq_1\sqrt{\left(\frac{2h}{a^2}-1\right)q_1^2+
\frac{\epsilon}{2a^e}h^2+h-g } +
\int dq_2\sqrt{\left(\frac{2h}{a^2}-1\right)q_2^2+
\frac{\epsilon}{2a^4}h^2+h+g },
\label{ex.9}
\end{eqnarray}
parametrized by two arbitrary parameters $h$ and $g$.
From this solution it is clearly seen that for $h<a^2/2$
the motion in the system is finite, in full accordance with
general reasonings given above.

Concluding the exposition of the classical case, let us
stress again the fact that in this case there is no
principal difference between the models associated with 
functions (\ref{ex.1}) and (\ref{ex.2}) with linear and
non-linear dependence on spectral parameters.
We have seen that in non-linear case the models turned out
to be more complicated than in the linear one, but they are
still ordinary classical models admitting quite standard
interpretation in terms of Hamilton -- Jacobi equation.

\subsection{Quantum case}

Let us now use functions (\ref{ex.1}) and (\ref{ex.2}) 
for building quantum versions of the models discussed in the
previous subsection. We start with multi-parameter spectral
equations associated with these functions and having the form:
\begin{eqnarray}
\left\{-\hbar^2\frac{\partial^2}{\partial q_1^2}+
q_1^2-\frac{\epsilon}{2a^4} h^2-
\left(\frac{2}{a^2}q_1^2 +1\right)h+g\right\}\Psi_1(q_1)=0
\label{ex.11}
\end{eqnarray}
and
\begin{eqnarray}
\left\{-\hbar^2\frac{\partial^2}{\partial q_2^2}+
q_2^2-\frac{\epsilon}{2a^4} h^2-
\left(\frac{2}{a^2}q_2^2 +1\right)h-g\right\}\Psi_2(q_1)=0
\label{ex.12}
\end{eqnarray}
It is not difficult to see that the system of these equations will have
discrete spectrum if we require the square integrability of
both functions $\Psi_1(q_1)$ and $\Psi_2(q_2)$. Indeed, let us rewrite
these equations in the form
\begin{eqnarray}
\left\{-\hbar^2\frac{\partial^2}{\partial q_1^2}+\omega^2
q_1^2\right\}\Psi_1(q_1)=e_1\Psi_1(q_1)
\label{ex.13}
\end{eqnarray}
and
\begin{eqnarray}
\left\{-\hbar^2\frac{\partial^2}{\partial q_2^2}+\omega^2
q_2^2\right\}\Psi_2(q_2)=e_2\Psi_2(q_2)
\label{ex.14}
\end{eqnarray}
with
\begin{eqnarray}
\omega=\sqrt{1-\frac{2h}{a^2}}
\label{ex.15}
\end{eqnarray}
and
\begin{eqnarray}
e_1=\frac{\epsilon}{2a^4}h^2+h-g,\qquad e_2=\frac{\epsilon}{2a^4}h^2+h+g.
\label{ex.16}
\end{eqnarray}
We see that formulas (\ref{ex.13}) and (\ref{ex.14}) look
as ordinary Schr\"odinger equations for simple harmonic oscillators,
and therefore, if the functions $\Psi_1(q_1)$ and $\Psi_2(q_2)$
are normalizable, have standard solutions
\begin{eqnarray}
\Psi_1(q_1)={\cal H}_n(\sqrt{\omega/\hbar}q_1)
\exp\left\{\frac{\omega q_1^2}{\hbar}\right\}, \qquad
\Psi_1(q_2)={\cal H}_m(\sqrt{\omega/\hbar}q_2)
\exp\left\{\frac{\omega q_2^2}{\hbar}\right\}
\label{ex.17}
\end{eqnarray}
and
\begin{eqnarray}
e_1=\omega\hbar(2n+1), \qquad e_2=\omega\hbar(2m+1),
\label{ex.18}
\end{eqnarray}
where by ${\cal H}_n$ and ${\cal H}_m$ we denoted the
ordinary Hermite polynomials.
Comparing formula (\ref{ex.18}) with (\ref{ex.15}) and (\ref{ex.16}),
we obtain the system of two equations
\begin{eqnarray}
\frac{\epsilon}{2a^4}h^2+h-g=(2n+1)\hbar\sqrt{1-\frac{2h}{a^2}},\qquad
\frac{\epsilon}{2a^4}h^2+h+g=(2m+1)\hbar\sqrt{1-\frac{2h}{a^2}}
\label{ex.19}
\end{eqnarray}
for $h$ and $g$. The sum of these equations gives a single
equation for $h$:
\begin{eqnarray}
\frac{\epsilon}{2a^4}h^2+h=(n+m+1)\hbar\sqrt{1-\frac{2h}{a^2}},
\label{ex.20}
\end{eqnarray}
which, obviously, has a discrete set of solutions. 

Consider particular cases of equation (\ref{ex.18}). Let
$a$ be large. Then the solution of this equation can be written
down immediately. It is
\begin{eqnarray}
h=(n+m+1)\hbar.
\label{ex.21}
\end{eqnarray}
We see that the spectrum of the parameter $h$ is unbounded.
Let now $a$ be finite and $\epsilon$ be small. Then
we obtain a quadratic equation for $h$. Solving it and
choosing that solution which is a continuous deformation of
(\ref{ex.21}), we obtain
\begin{eqnarray}
h=\frac{(n+m+1)^2\hbar^2}{a^2}\left\{-1
+\sqrt{1+\frac{a^4}{(n+m+1)^2\hbar^2}}\right\}.
\label{ex.22}
\end{eqnarray}
In this case the spectrum of the parameter $h$ is bounded
by $h_{max}=a^2/2$. If the numbers $n,m$ are large, then the
distances between neighbouring spectral points are small,
so that the point $h_{max}=a^2/2$ is the accumulation point of
the spectrum.
The general case of arbitrary $\epsilon$  and $a$ 
can be considered analogously. The explicit solution of
equation (\ref{ex.20}) is very complicated, because (\ref{ex.20})
is now fourth order algebraic equation. However, the
qualitative behaviour of the spectrum is the same as in the
last case. As before, the spectrum is bounded by $h_{max}=a^2/2$ and
this is its accumulation point. 

Thus, we have demonstrated the normalizability of functions
$\Psi_1(q_1)$ and $\Psi_2(q_2)$ as well as the discreteness
and infiniteness of the spectrum of equations 
(\ref{ex.11}) and (\ref{ex.12}) in the interval $0<h<a^2/2$.
Analogously, it can be shown that for $h>a^2/2$ the functions
$\Psi_1(q_1)$ and $\Psi_2(q_2)$ are non-normalizable and
the spectrum of multi-parameter spectral equations
(\ref{ex.11}) and (\ref{ex.12}) is continuous, in full
accordance with the classical case.

Let us now try to understand whether the spectral values of the
parameter $h$ can be considered as the eigenvalues of a certain
two-dimensional operator in Hilbert space.
In order to do this, it is sufficient to follow the
general prescriptions of section \ref{qu}. Let us multiply
the first equation (\ref{ex.11}) by $\Psi_2(q_2)$, and the
second one, (\ref{ex.12}), by $\Psi_1(q_1)$. Introducing the
function $\Psi({\bf q})=\Psi_1(q_1)\Psi_2(q_2)$ and adding
the equations (\ref{ex.11}) and (\ref{ex.12}), we get a single equation for
$\Psi({\bf q})$, 
\begin{eqnarray}
\left\{  -\hbar^2\Delta + {\bf q}^2- \frac{\epsilon}{a^4} h^2-2
\left(\frac{1}{a^2} {\bf q}^2 +1 \right)h \right\}\Psi({\bf q})=0,
\label{ex.23}
\end{eqnarray}
containing a single spectral parameter $h$. Consider first
the particular cases of this equation.

Let $a$ be large. Then the terms proportional to $h^2$ and
$h{\bf q}^2$ in (\ref{ex.23})
vanish and we obtain the ordinary Schr\"odinger equation
for a two-dimensional harmonic oscillator 
\begin{eqnarray}
\frac{1}{2}\left\{  -\hbar^2\Delta + {\bf q}^2
\right\}\Psi({\bf q})=h\Psi({\bf q}).
\label{ex.24}
\end{eqnarray}
Let now $a$ be finite and $\epsilon$ small. Then the only term
proportional to $h^2$ in (\ref{ex.23}) vanishes and we
obtain again a linear spectral equation but with the weight function
${{\bf q}}^2/a^2+1$. In order to reduce the hamiltonian of this equation to
the hermitean form, we can redefine the wavefunctions as
$\Psi({\bf q})\rightarrow ({{\bf q}}^2/a^2+1)^{1/2}\Psi({\bf q})$,
after which the equation becomes
\begin{eqnarray}
\frac{a^2}{2}\frac{1}{\sqrt{{{\bf q}}^2+a^2}}
\left\{  -\hbar^2\Delta + {\bf q}^2 \right\}
\frac{1}{\sqrt{{{\bf q}}^2+a^2}}\Psi({\bf q})=h\Psi({\bf q}).
\label{ex.25}
\end{eqnarray}
Let us now consider the general case of arbitrary $a$ and $\epsilon$.
Now the spectral parameter $h$
does not enter in equation (\ref{ex.23}) linearly and thus, cannot
be considered as an eigenvalue of some linear operator.
The only thing what we can do in this case, is to solve
the equation (\ref{ex.23}) with respect to $h$, and then, 
multiplying the result by $\Psi({\bf q})$ and exchanging
the left- and right-hand sides, write the nonlinear differential
equation
\begin{eqnarray}
\frac{a^2}{\epsilon^2}({\bf q}^2+a^2)\left\{-1
+\sqrt{1+\frac{\epsilon^2\left(-\hbar^2\Delta\Psi({\bf q})+
{\bf q}^2\Psi({\bf q})\right)}{(a^2+{\bf q}^2)^2\Psi({\bf q})}}\right\}
\Psi({\bf q})=h\Psi({\bf q})
\label{ex.26}
\end{eqnarray}
Note that the ordinary (linear) quantization of the classical
system with hamiltonian $H$ leads to the linear
Schr\"odinger equation of the form
\begin{eqnarray}
\frac{a^2}{\epsilon^2}({\bf q}^2+a^2)\left\{-1
+\sqrt{1+\frac{\epsilon^2\left(-\hbar^2\Delta+{\bf q}^2\right)}
{(a^2+{\bf q}^2)^2}}\right\}\Psi({\bf q})=h\Psi({\bf q})
\label{ex.27}
\end{eqnarray}
with all necessary permutations of non-commuting operators
guaranteeing the hermitian symmetry of the corresponding hamiltonian.
Expanding equations (\ref{ex.26}) and (\ref{ex.27}) in 
$\hbar^2$, it is easy to see that the first two terms of these
expansions coincide. The difference (the non-linearity) appears
only in the third term and is of order $\hbar^2$.
This non-linearity is of the type $(\Delta\Psi/\Psi)^2\Psi$ and is
very similar to non-linearities appearing in Doebner -- Goldin 
quantization scheme. 

\section{Conclusion}

Of course, the example we discussed here, is
rather artifical and hardly has some relation to reality (even if 
one belives that quantum world is non-linear), but it clearly
demonstrates the idea lying in the ground of our approach. At any rate,
irrespective of the physical reasonability of models obtainable by
means of our non-linear quantization procedure, the method which we
proposed provides a constructive way for extending the notion of
quantum integrability to non-linear spectral problems and gives
a practical tool for buiding completely integrable non-linear spectral
equations in Hilbert space. 

An interesting application
of this method would be construction of non-linear versions
of quantum Gaudin models associated with higher Lie
algebras. As it was noted in ref. \cite{ric}, there exist
multi-parameter spectral equations whose solutions exactly coincide
with Bethe ansatz solutions for the Gaudin models. Some of these
equations depend on spectral parameters linearly. The 
application of quantum IMSV to such equations leads to ordinary
(linear) Gaudin models (the simplest $sl(2)$ case of such a
transformation was considered in detail in ref. \cite{ushbook}).
However, if the rank of a Lie algebra is suffuciently high,
then, along with the "linear" multi-parameter spectral equations,
there are equations
in which the spectral parameters enter non-linearly. The application
of quantum IMSV to such equations should lead to completely
integrable models having the same spectra as the ordinary
Gaudin models but realized by non-linear operators in
Hilbert space. An explicit construction of such models is
an interesting mathematical problem and we
hope to consider it in one of the fortcoming publications.

\section{Acknowledgements}

One of us (AU) thanks the staff of Theoretical
Physics Department of the University of Parma for kind
hospitality.


\begin{thebibliography}{99}
\bibitem{dob1} H.-D. Doebner and G.A. Goldin, Phys. Lett. A
{\bf 162} 397 (1992) 
\bibitem{gol} G.A. Goldin, R. Menikoff and D.H. Sharp,
Phys.Rev.Lett. {\bf 51}, 2246 (1983)
\bibitem{ang} B. Angermann, H.-D. Doebner and J. Tolar,
in: Lecture Notes in Mathematics, vol. 1037,  Nonlinear
partial differential operators and quantization procedures,
(Springer: Berlin, 1983), p. 171
\bibitem{bial} I. Bialynicky-Birula and J. Mycielski, Ann. Phys.
{\bf 100}, 62 (1976)
\bibitem{kib} T. Kibble, Comm. Math. Phys. {\bf 64}, 73 (1978) 
\bibitem{schu} D. Schuch, K.M. Chung and H. Hartman, J. Math. Phys.
{\bf 25}, 3086 (1984)
\bibitem{wein} S. Weinberg, Ann. Phys. {\bf 194}, 336 (1989)
\bibitem{skl} E.K. Sklyanin, Preprint of Helsinki University
HU-TFT-91-51, Helsinki (1991), (see also hep-th/9211111)
\bibitem{dobush} H.-D. Doebner and A.G. Ushveridze, Integrable and
algebraically solvable systems, in: H.-D. Doebner, V.K.
Dobrev and A.G. Ushveridze (Eds), Generalized symmetries in
physics (World Scientific: Singapore, 1994) p. 225 
\bibitem{skl2} E.K. Sklyanin, Preprint of Cambridge University
NI-92013, Cambridge (1992)
\bibitem{ushleb} A.G. Ushveridze, Sov. J. Part. Nucl. {\bf
20}, 504 (1989)
\bibitem{ushbook} A.G. Ushveridze, Quasi-exactly-solvable models in
quantum mechanics (IOP Publishing: Bristol, 1994)
\bibitem{ric} A.G. Ushveridze, Generalized Gaudin Models
and Riccatians, in: Complex Analysis and its Applications
(Banach Center Publications: Warsaw, 1995)

\end{thebibliography}
\end{document}